\documentclass[english,12pt]{article}
\usepackage{amssymb}
\usepackage{epsfig}
\usepackage{babel}
\begin{document}
\centerline{\bf Properties and application of the form $A\cdot exp(-(x-c)^2/(a(x-c)+2b^2))$}
\centerline{\bf for investigation of ultra high energy cascades}
\vskip12pt

\centerline{A.A. Kirillov*, I.A. Kirillov}

\it
{D.V. Skobeltsin Institute of Nuclear Physics, M.V. Lomonosov Moscow State University,

119992, Moscow, Russia}
\rm
\vskip12pt
\vskip12pt

\begin{abstract}

The form $A\cdot exp(-(x-c)^2 /(a(x-c)+2b^2))$ is an asymmetric distribution intermediate between the normal and exponential distributions. Some specific properties of the form are presented and methods of approximation are offered. Appropriate formulae and table are presented. The practical problems of approximation by the form are discussed with connection to the quality of original data. Application of the methods is illustrated by using in the problem of calculating and studying distribution function of maximum ultra high energy atmospheric showers. Relationship with exponential and normal distributions makes usage of the form to be effective in practice.
\end{abstract}



\vskip12pt
\vskip12pt

*Corresponding author Alexander Kirillov.
Tel.: +7-495-939-2437; fax: +7-495 939 3553

e-mail address: krl@dec1.sinp.msu.ru (A.A. Kirillov).

\newpage

\section{Introduction}

There are many phenomena that have: 1) bounds on their characteristics, 2) large number of relatively small factors composing the value of these characteristics, 3) the characteristic development rate is proportional to their values. To describe these phenomena, the main features of the function under consideration should be the following: 1) the domain of definition is the half-line, 2) the peak region is similar to the peak of the normal distribution and 3) the tail falls exponentially.

These kind phenomena are in high energy cascades. To investigate the giant energy cascades (about $\approx 4\cdot 10^{19}$ eV - the conditional threshold of the relic cut-off [1]), the approximation of the total number of particles of the individual cascades, $N(t)$, at depth $t$ of the atmosphere, the function $N_m exp(-(t-t_m)^2 /(a(t-t_m)+2b^2))$ has been successfully used [2]. The parameters of the function have an obvious intuitive interpretation in this case: ($t_m$, $N_m$) is the cascade maximum position, $a$ is the asymmetry, $b$ is the breadth.
The physical phenomenon can be seen to have the above mentioned three properties as definition properties as does the functional form:
\begin{equation}
f_{en}(x)=A(a,b)exp(-(x-c)^2/(a(x-c)+2b^2)) \ \ a\ge 0,\ x\ge c-2b^2/a \ \ \mbox{or} \ \ a<0,\ x<c-2b^2/a.
\end{equation}
The giant or ultra high energy cascades (UHE $\Leftrightarrow E_0\ge 10^{18}$ eV) are rare events: 1 event per $\approx 1$ km$^2 $ sr in a year at $E_0\ge 5\cdot 10^{18}$ eV; 1 event per $\approx 1$ km$^2 $ sr in a century at $E_0\ge 1\cdot 10^{20}$ eV. Each event is examined individually. The total number of the detected cascades of energy $E_0\ge 5\cdot10^{18}$eV is about 15000 and those of energy $E_0\ge 1\cdot 10^{20}$ eV is about 200 to date (end 2007 yr.). The must fruitful experiment with the integrated exposure mounts up to about 5165 km$^2$ sr yr (Pierre Auger Observatory 01.01 2004 $\div$ 28.02 2007) has 11853 events $E_0 > 3\cdot 10^{18}$ eV and only 2 events $E_0 > 10^{20}$eV. The situation due to not technical specifies experimental features only, but shows flux suppression at the highest energies. The statistical and systematic uncertainness in the energy scale are of the order of 6\% and 22\%, respectively [3]. Technical difficulties cause registration of a part of flux and lower quality of the processing data.

Simulation of UHE cascades is multi-step-by-step process requiring large resources, so it is carried out by a hybrid method with plenty of internal assumptions and averaging [4]. A typical claimed accuracy of simulation results is $\approx$ 5\% (felt by some to be an optimistic estimation). The smallest change in the simulation, made by the same author, changes results by approximately 10\%. At each stage of the multistage process, a method of processing both experimental and model data is chosen on the basis of the quality of this stage's input data.

Form (1) has clear exponential asymptotic, parameterization of the $f_{en}(x)$ is visual and has correct geometrical interpretation (next section (15), (16)) besides the intuitive interpretation. Since the form is very simple, it may be used as basic distribution to approximate poor quality data. Practical usage of the refined apparatus of distribution functions to describe the phenomenon requires paying attention to the initial data quality. Descriptive ability of $f_{en}(x)$ is rather wide and generalizes the abilities of exponential and normal distribution (see fig. 2), therefore $f_{en}(x)$ may has wide applications. UHE cascades maximum position generated by protons in atmosphere was investigated using the $f_{en}(x)$ as probability density [5].

Properties of the asymmetry and of the transformation to normal distribution give ability to treat $f_{en}(x)$ for approximation of sum of finite number, $n$, random values with accuracy $1/n$ (see next section (22) \ -- \ (26)). On analogy of the normal distribution, $f_{en}(x)$ can be interpreted as distribution of the measurement error at finite number of factors, $n$, which affect on asymmetric elementary measuring or as approximation sum of finite number of asymmetric random variables.

The present paper deals with the methods of approximation by this form of data taking into account their completeness, quality and the type of the presentation independently of whether they can have probabilistic interpretation or not. Approximation of any distribution by formulae simple and comfortable for both analytical and numerical methods is an ordinary task, often appearing when simulating and processing data. The exponential and normal distributions are often used not only because they represent distributions resulting from a physical process but also because they are, in our opinion, comfortable to use. One glance at a graph or histogram is enough to estimate their distribution parameters. Form (1) inherits the suitability.

Experimental or simulation results are usually represented by the mean value and variance only. Often, however, the data, considered as random variables are asymmetric. Therefore, representation of data by the mean value and variance only is insufficient. The next step is representation of the variance as a sum of left and right parts, and form (1) is convenient for the representation. The present article intends to show that the asymmetry can be taken into account even in a case of poor quality data.

At the beginning of the article, the properties of form (1) are presented as the simplest case between the exponential and normal distributions. Then methods of approximation are developed with taking into account the initial data quality and, finally, the methods are used to study UHE cascades. These last sections (thought may be interesting in itself) illustrate use of the methods. In our opinion, the elaboration has general character and may be useful in other fields.

\section{Some properties of form $A(a,b)exp(-(x-c)^2/(a(x-c)+2b^2))$}

The parameterization is convenient for interpretation of the parameters. For standard support $x \in (0, \infty)$ the form is $B(s,q)exp(-(sx+q/x)/2)$, where $s \ge 0$, $q \ge 0$.
The distribution $f_{en}(x)$ is infinitely-divisible [6, b] and belongs to the family of Generalized inverse Gaussian distribution (investigated by B. Jorgensen [7]). The probability density of the family is
$$f_{GiG}(x)=\frac{(s/q)^{p/2}}{2K_p(\sqrt{sq})}x^{p-1}e^{-(sx+q/x)/2},$$
where $x>0$, $K_p$ is a modified Bessel function of the third kind, with index $p<\pm \infty$, $s>0$, and $q>0$.

The distribution $f_{en}(x)$ is a special case [6, a] of Hyperbolic distributions: its hyperbola has vertical asymptote. The special case generates the family Hyperbolic distributions (introduced in and investigated by O.E. Barndorff-Nielsen [6]) by normal variance-mixing. The probability density of the family is:
$$f_H (x)=\frac{\sqrt{s/q}}{2\sqrt{\beta^2+s} \ K_1(qs)}e^{-\sqrt{((x-\mu)^2+q)(\beta ^2+s)}+(x-\mu)\beta},$$
where $x \in (-\infty, +\infty)$, $\mu, \beta$, $s>0$ and $q>0$ are parameters. The family is normal variance-mean mixture $Y=\mu+\beta V+\sigma \sqrt{V}X$ at $\sigma=1$, $\mu$ and $\beta$ are free parameters. Here random variables $X$ and $V$ are independent, $X$ is normal distributed with mean zero and variance one, and $V$ has probability density $f_{en}(x)$ on support $x \in (0, \infty)$.

Formulae for the general characteristics (normalization, moments, Laplace transform, Fourier transform) of the distribution $f_{en}(x)$ may be obtained with help of formulae 3.471.9 of Gradsteyn and Ryzhik [8]:
\begin{equation}
\int _{0}^{\infty}x^{p-1}exp(-(sx+q/x)/2)dx=2(s/q)^{p/2}K_p(\sqrt{sq}) \ \ \ Re(s)>0 \ \ \ Re(q)>0,
\end{equation}
where $K_p$ is a modified Bessel function of the third kind, with index $p$.

In particular, for normalization factor of $f_{en}(x)$ we have
\begin{equation}
A(a,b)=[a(2b/a)^2 exp((2b/a)^2)K_1((2b/a)^2)]^{-1}.
\end{equation}
Asymptotic expansions [8, 9] for $K_1(z)$
$$K_1(z)=\frac{1}{z}+\frac{z}{2}ln\frac{z}{2}+o(z^2) \ \ \ \ z\to 0,$$
\begin{equation}
K_1(z)=\sqrt{\frac{\pi}{2z}}e^{-z}(1+\frac{3}{8z}-\frac{15}{128z^2}+\ .\ .\ .) \ \ \ \ z\to \infty
\end{equation}
show that the $f_{en}(x)$ tends to the exponential distribution, $f_e(x)$, as $z \rightarrow 0 $
and to the normal one, $f_n(x)$, as $z \rightarrow \infty $.
distributions.

At scaling $x\Rightarrow x/\alpha$ with saving equality
$\frac{1}{\alpha}f(\frac{x}{\alpha},a)=f(x,a^{\prime})$ the parameters of the $f_e (x)$ and $f_n (x)$ are transformed $a^{\prime}=a\cdot\alpha$. The parameters $a$ and $b$ of the $f_{en}(x)$ are transformed in the same way. Thus, $a/b=a^\prime /b^\prime$ therefore this ratio is scaling invariant and can be considered as the essential parameter of $f_{en}(x)$. It is convenient to define the scaling-invariant parameter
\begin{equation}
z=\left( 2\frac{b}{a}\right) ^2,
\end{equation}
take it as a parameter of type of the distribution $f_{en}(x)$
and write the normalizing factor (3) in the form:
$$A(a,b)=[a\cdot z\cdot exp(z)K_1(z)]^{-1}.$$
Obviously, the parameter $z$ may be treated as a measure of asymmetry (skewness) of the $f_{en}(x)$, but at data processing it is convenient to use another scaling-invariant function also.

We define $\sigma_l^2$ and $\sigma_r^2$ as left and right
parts of the variance $\sigma^2$ (above the mean value) by the formulae:
\begin{equation}
\tilde X=\int ^{\infty}_{c-2b^2 /a}{x f(x)dx},
\end{equation}
\begin{equation}
\sigma ^2=\int\limits _{-\infty} ^{+\infty} (\tilde X-x)^2 f(x)dx=\sigma _l^2+\sigma _r^2=\int\limits _{-\infty}^{\tilde X}(\tilde X-x)^2 f(x)dx+\int\limits _{\tilde X}^{+\infty}(\tilde X-x)^2 f(x)dx.
\end{equation}
The ratio
\begin{equation}
R^2 =\sigma _l^2/\sigma _r^2
\end{equation}
is invariant under scaling of $x$ and therefore $R^2$ depends on $z$ only, and also may be taken for measuring of asymmetry. For exponential distribution $R^2=e/2-1$. Since $R^2$ for $f_{en}(x)$ is monotonic function bounded between its values for $f_e(x)$ and $f_n(x)$, we have
\begin{equation}
e/2 -1=0.359..\le R^2\le 1,
\end{equation}
for $a \ge 0$. Fig. 1 shows the measure of asymmetry, $1-\sigma _l^2/\sigma _r^2$, of the $f_{en}(x)$ from exponent to normal distributions.

One can see that $f_{en}(x)$ is a unimodal, bell-shaped, right-asymmetric ($a>0$) distribution. To illustrate the descriptive ability of $f_{en}(x)$, fig. 2 shows the distributions at $c$=0, $b$=1 and variable $a $=0, 1, 5, 10. One can see that $f_{en}(x)$ changes from $f_n(x)$ towards $f_e(x)$.
The dotted line in fig. 2 is the easily recognized standard form of the normal distribution at $a$=0. An examination of the figure shows the distributions at $a \ne 0$ have supports on half-lines, with the left ends located on the hyperbola $a=-2/x$ according to the conditions in (1). The maximum density (equal to the value of normalizing factor $A(a,b)$) decreases with increasing $a$. For $a$ =5 the value A(5,1) is singled out in the figure. The distance, $\tilde x$, between the mode, $c$, and the mean value, $\tilde X$,
\begin{equation}
\tilde X=\tilde x + c
\end{equation}
increases with $a$ gradually because the distributions are steepening on the left side and flattening in the right side. It can easily be imagined how the form of the distribution is changed during scaling and see that no one of its can be obtained from other by scaling.

Thus, probability density $f_{en}(x)$ defines a one-parameter set of distributions with rather wide descriptive ability.

Note that the mode is a more natural characteristic of the distribution $f_{en}(x)$ than the mathematical expectation, whose distance from the mode characterizes asymmetry. Such kind of asymmetry is known as Pearson's skewness: $(\tilde X-c)/\sigma$, which may be obtained from lines 3 and 4 of table 1 (see below).

It is interesting to study the width of (1), $W$, with its left part, $W_l$, and right part, $W_r$, at the mode: $W=W_l+W_r$ (see fig. 3).
The width of distribution is used quite often. For example, an error bar graph is usually used to characterize the accuracy of measurement results. The $-\sigma$ and $+\sigma$ values represent a normal distribution width at the $e^{-1/2}$ level, i.e. at the level $d^{-1}= e^{1/2}$ where the probability density is less by a factor of $e^{1/2}$ than at its maximum. In some cases the width of distribution has a physical interpretation and measures a physical quantity, for example, a Cherenkov light lateral distribution, or the lifetime of a resonance.
The width is less fluctuating characteristic then other and is used as robust characteristic. For UHE events heavy nuclei are expected to produce smaller shower-to-shower fluctuations then protons. Therefore the width of distributions of its characteristic is sensitive parameter. These properties are useful and are used wide [10.11.12].

These $W$, $W_l$, $W_r$ at the level $d$ is defined by solutions of the equation:
$$A(a,b)exp(-x^{2}/(ax+2b^2))=A(a,b)/d=A(a,b)exp(-\ln d).$$
One can see from the equation $x^2 -x\cdot a \ln d-2b^2 \ln d=0$ (see also fig. 3) that

1. The width of the distribution is determined by the discriminant of the equation:
\begin{equation}
W=\sqrt{a^2 (\ln d)^2 +8b^2 \ln d}.
\end{equation}

2. The parameter $a$ is determined by the difference of the right and the left parts of the distribution width (at the mode $c=0$) i.e.
\begin{equation}
a=(W_r -W_l )/\ln d.
\end{equation}

3. The parameter $b$ is determined by the product of the left and right parts of the width
\begin{equation}
2b^2 = W_lW_r/\ln d \, \mbox{ or }\, b=( W_lW_r/(2\ln d))^{1/2}.
\end{equation}

For future convenience rewrite (11)--(13) for the level $d=e$:
\begin{equation}
W=\sqrt{a^2 +8b^2},
\end{equation}
\begin{equation}
a = W_r-W_l,
\end{equation}
\begin{equation}
2b^2 =W_lW_r \ \ \mbox{or} \ \ b=(W_lW_r /2)^{1/2}.
\end{equation}

Formulae (15), (16) have evident geometrical interpretations (see also fig. 3) which induce a physical interpretation of parameters in particular cases.

Note the relationship between the widths and the parameter $z$ (5) defining the distribution form
$$ \frac{W_lW_r}{(W_r-W_l)^2} 2\ln d =z \ \mbox{at any level d}.$$

Obviously, that the width at one level is not sufficient to determine $f_{en}(x)$, but adding the ratio parts of the width (divided by the mode or the mean) is sufficient, as it is shown by (11) -- (13).

One can see some analogy of ways to describe the distribution function in terms of width and variance, because both of it's describe a breadth, but width is function of the level $d$. It is possible to create approximation of UHE cascade profile in term of width [13].
 Mistakes caused by the habit of believing that the width of distribution is defined by the standard deviation should be avoided. For example, if it is assumed that $W=2\sigma$ (at the level $d^{-1}=e^{-1/2}$ ), i.e. if the relationship between the width and the variance is assumed to be the same as for the normal distribution, one obtains a formula for the variance: $\sigma ^2=b^2+a^2/16$, that is true at $a=0$ only.

The same as for normal distribution often it is needed cumulative form of the $f_{en}(x)$ distribution.

The function
\begin{equation}
F_{en}(X,a,b,c)=\int _{-2b^{2}/a+c}^{X} {\frac{exp(-(x-c)^2 /(a(x-c)+2b^{2}))}{a(2b/a)^{2}exp((2b/a)^2)K_l((2b/a)^2)}}dx
\end{equation}
is the integral form of one-parametrical distributions $f_{en}(x)$. Since, for practical using, it is convenient to have (17) together with the table binding the values of parameters $a, b, c$ with the basic characteristics of the distribution (the normalizing factor, the mathematical expectation, the variance, and skewness ), the standard form $f_{en}(x)$ is taken in form:
\begin{equation}
F_{en}(y,z) =\int _{-z/2}^{y}{\frac{exp(-x^2/(x+z/2))}{z\cdot exp(z)\cdot K_l(z)}dx}
\end{equation}
corresponding to $a=1, c=0$. Thus,
\begin{equation}
F_{en}(X,a,b,c)=F_{en}((X\mp c)/(\pm a),(2b/a)^2)=F_{en}(y,z),
\end{equation}
where the sign at the parameter $c$ is opposite to the sign of the parameter $a$ value.

So, for one-parametrical distribution $f_{en}(x)$ we have table 1 with lines: $z$, $A\vert a \vert$, $\tilde x /a$, $\sigma ^2 /a^{2}-1$, $\sigma ^2_l /\sigma ^2_r$, $\vec y$, where the vector designates the lines $F_{en}(\vec y, z)$ (18) for the $y$ values, listed in the first column. After calculating $z=(2b/a)^2$; the values of the normalizing factor, mathematical expectation, variance, and the ratio of components of the variance can be obtained using the values of parameters $(a, b, c)$ from the top 5 lines; and conversely, the values of mean, the variance and the ratio of components of variance can be used to obtain the values of $A(a,b), a, b, c$ (see below in section 3 the description of method A and the numerical example in section 5, footnote 3). Dealing with the cumulative form (the distribution function) is defined by formula (19) determining $F_{en}(X,a,b,c):$ it is necessary to find in the table the value of $F_{en}(y,z)$, where $y=(X\mp c)/(\pm a), z=(2b/a)^2$. The numerical example see in section 5, footnote 4.

Let us present the list of the basic properties of the $f_{en}(x)$ as bridge between the exponential and the normal distributions.

\vskip 0.5cm
Distributions $ \hspace {18mm} \frac{exp(-(x-c))/a)} {a}$ \ \ \ \ \ \ \ $\frac{exp(-(x-c)^2 /(a(x-c)+2b^2))} {a(2b/a)^2 exp((2b/a)^2)K_l ((2b/a)^2)}$ \ \ \ \ \ \ \ \ \ $\frac{exp(-(x-c)^2 /(2\sigma ^2))} {\sqrt{2\pi}\sigma}$

support\hspace {30mm} $c\le x \le\infty$ \ \ \ \ \ \ \ \ \ \ $-2b^2 /a+c\le x\le\infty$ \ \ \ \ \ \ \ \ \ \ \ \  $\infty\le x\le +\infty$

Symmetry \hspace {30mm} \ -- \hspace {35mm} \ -- \hspace {47mm} \ +

Interpret.
of parameters \hspace {1mm} \ +, + \hspace {25mm} \ +, \ +, \ + \hspace {35mm} \ +, \ +

Scaling \hspace {35mm} \ + \hspace {32mm} \ +, \ + \hspace {42mm} \ +

Unscaling
parameter \hspace {13mm} -- \hspace {25mm} \ $b/a \ \ [z=(2b/a)^2]$ \hspace {28mm} \ --

Mode \hspace {40mm}\ $c$ \hspace {35mm} \ $c$ \hspace {48mm} \ $c$

Characteristic
Function \hspace {5mm} $\frac{e^{itc}}{1-ita}$ \hspace {20mm} \ $\frac{e^{it(c\frac {2b^2}{a})}}{\sqrt{1-ita}} \frac{K_1(z\sqrt{1-ita)}}{K_1(z)}$ \hspace {25mm} $e^{-\frac{t^2\sigma ^2}{2}+itc}$\

Mathematical

expectation \ \ $\tilde X $ \hspace {20mm}$c+a$ \hspace {15mm} $c+\frac{2b^2}{a}(K_2(z)/K_1(z)-1)$ \hspace {25mm} $c$

Variance \hspace {18mm} $a^2=(\tilde X-c)^2$ \hspace {10mm} $4b^2[(\frac{K_2(z)}{K_1(z)})^2(-\frac{Z}{4})+\frac{K_2(z)}{K_1)z)}+\frac{z}{4}]$ \hspace {15mm} $\sigma ^2=b^2$

Third central

moment \ $\mu_3$ \ \ \ \ \ \ \ \ $2a^3$ \hspace {2mm} $a4b^2[(\frac{K_3(z)}{K_1(z)})^3 (\frac{z^2}{4})+(\frac{K_2(z)}{K_1(z)})^2(-\frac{3}{2}z)+\frac{K_2(z)}{K_1(z)}(-\frac{z^2}{4}+3)+\frac{3}{4}z^2]$ \hspace {5mm} $0$

Ratio $\sigma _l^2/\sigma _r^2, \ (a>0)$ \hspace {8mm} $e/2-1$ \hspace {17mm} $e/2 -1\le \sigma _l/\sigma _r\le 1$ \hspace {30mm} 1

Width at level $e^{-1}$ \ \ \ \ \ \ \ \ \ \ \ $a$ \ \ \ \ \ \ \ \ \ \ \ $\sqrt{a^2 +8b^2}$ \ \ $[W_{l,r}=W/2\mp a/2]$ \ \ \ \ \ \ \ \ \ \ \ \ \ \ \ $2\sqrt{2}\sigma$
\vskip 0.5cm
One can see how properties of the normal or the exponential distributions for the intermediate case under consideration are generalized, transforming to appropriate relations at $a\rightarrow 0$ or $b\rightarrow 0$. The $f_{en}(x)$ enables description of some asymmetric distributions. The possibility is furnished by parameter $a$ but, in general, it is incorrect to think of this as a skewness parameter, because its value is not scaling-invariant. Any monotonic function depending on $a/b$ only may be taken as a measure of asymmetry (skewness) for $f_{en}(x)$, for example $1/z$ defined by (5) or $1-R^2$ defined by (8).

Note one more property of $f_{en}(x)$ caused by transformation of $f_{en}(x)$ to normal distribution.

Since under weak conditions, sum of $n$ symmetric random variables is approximated by the normal distribution with an accuracy $\sim 1/n$, but asymmetric ones with accuracy $\sim 1/\sqrt{n}$, that it is possible to find such asymmetric distribution which approximates the asymmetric sum with an accuracy $\sim 1/n$. One of the such kind distribution is $f_{en}(x)$.

Indeed, let $f(x)_{\Sigma}$ be the probability density of
$\chi=\sum_{i=1}^n \frac{\xi_i-m}{\sigma \sqrt{n}},$  where $\xi_i$ are the independent random variables with a common distribution which mean value is $m$ and standard is $\sigma$.
Write Edgeworth expansion [14] for the $f(x)_{\Sigma}$ including term $R_{4n}=O(n^{-1})$:
\begin{equation}
f(x)_{\Sigma}=\varphi(x)-n^{-
\frac{1}{2}}\frac{1}{3!}\gamma_1\varphi^{(3)}(x)+(\frac{1}{4!}\gamma_2\varphi^{(4)}(x)+\frac{10}{6!}\gamma_1^2\varphi^{(6)}(x))(n^{-1})-...,
\end{equation}
where $\varphi^{(k)}(x)$ is $k$-th derivative of $\varphi(x)=\frac{1}{\sqrt{2\pi}}e^{-\frac{x^2}{2}},$ $\gamma_1=\mu_3/\sigma^3$ and $\gamma_2=\mu_4/\sigma^4-3$ are the skewness and excess of the $\xi_i$ distribution. Reminder that $\varphi^{(k)}(x)=(-1)^kH_k(x)\varphi(x),$ and $H_k(x)$ is $k$ degree Hermit polynomial.
Using $f_{en}(x)$ for approximation of $f(x)_{\Sigma}$, we have the discrepancy
\begin{equation}
\Delta=\varphi(x)-n^{-\frac{1}{2}}\frac{1}{3!}\gamma_1\varphi^{(3)}(x)-A(a,b)exp(-(x-c)^2/(a(x-c)+2b^2))+O(n^{-1}).
\end{equation}
We will take such parameters $a, \ b, \ c$ that term with $n^{-\frac {1}{2}}$ will be covered by $f_{en}(x)$. From (3) and (4) we have $A(a,b)=\frac{1}{\sqrt{2\pi}b}+O(a^2)$.
Dividing (21) by $\varphi(x)$ and using series for obtained exponent one can has:
$$(\Delta-O(n^{-1}))/\varphi(x)=1+n^{-\frac{1}{2}}\frac{1}{3!}\gamma_1(x^3-3x)-\frac{1}{b}(1-\frac{(x-c)^2}{a(x-c)+2b^2}+\frac{x^2}{2})+O(a^2)+O(c^2)$$
$$=n^{-\frac{1}{2}}\frac{1}{3!}\gamma_1(x^3-3x)-\frac{1}{b}+\frac{2x^2-4xc+2c^2-x^2a(x-c)-x^22b^2}{2b(a(x-c)+2b^2)}+O(a^2)+O(c^2)$$
$$=n^{-\frac{1}{2}}\frac{1}{3!}\gamma_1(x^3-3x)-\frac{ax^3}{2b(a(x-c)+2b^2)}- \frac{4cx}{2b(a(x-c)+2b^2)}$$
\begin{equation}
-\frac{(2b^2-2-ac)x^2}{2b(a(x-c)+2b^2)}+\frac{2c^2}{2b(a(x-c)+2b^2)}-\frac{1}{b}+1+O(a^2)+O(c^2).
\end{equation}
The equation shows the needed values of the parameters $a, \ b, \ c$:
\begin{equation}
a=\frac{2}{3}\gamma_1n^{-\frac{1}{2}}+O(n^{-1}),\ \ \ b=(1+O(n^{-1})^{\frac{1}{2}},\ \ \ c=-\frac{1}{2}\gamma_1n^{-\frac{1}{2}}+O(n^{-1}).
\end{equation}
Then by direct substitution one can be convinced that
$\Delta/\varphi(x)=O(n^{-1}).$ That is
$$f(x)_{\Sigma}=f_{en}(x)+\Delta+R_{4n}= f_{en}(x)+\varphi(x) O(n^{-1})+ R_{4n}=f_{en}(x)+O(n^{-1}).$$

According to (23), the parameters $a, \ b, \ c$ become to depend on $n^{-1/2}\gamma_1$, therefore the support (1) of the $f_{en}(x)$ depends on $n^{-1/2}\gamma_1$: $x\ge c-2b^2/a=-3n^{1/2}\gamma_1^{-1}-n^{-1/2}\gamma_1/2$. Out of the region we have: $0<f(x)_{\Sigma}\le O(n^{-1})$ and we define $f_{en}(x)=0$. The problem because of negative values of $H_3(x)$ [14, 1946; 6, 1989] does not arise.

Thus, the $f_{en}(x)$ uniformly approximates $f(x)_{\Sigma}$ for $-\infty \le x \le \infty $ with accuracy $O(n^{-1})$.

If $\xi_i$ have not common distribution [14, 1937], the $\gamma_1$ and $\gamma_2$ must be chosen by $$\gamma_1^{\prime}=\sum_{i-1}^n\mu_{3i}/(\sum_{i-1}^n\sigma_i^2)^{3/2}\ \ \mbox{and}\ \ \gamma_2^{\prime}=(\sum_{i-1}^n\mu_{4i}-3\sum_{i-1}^n\sigma_i^4)/(n^{-1}(\sum_{i-1}^n\sigma_i^2)^2),$$
where $\mu_{3i}$ and $\mu_{4i}$ are the third and the forth central moments of $\xi_i$, respectively.

The property has various consequences and seems to be useful for practical use.

As one can see, (23) can be defined more exactly. For example, one can obtain the $f_{en}(x)$, which three first moments coincide with the same of the $f(x)_{\Sigma}$: $0, \ \ 1, \ \ n^{-1/2}\gamma _1$. From (2) and (4) one can obtain the moments of the $f_{en}(x)$ in form:
$$m(f_{en}(x))=c+\frac{3}{4}a+\frac{3}{2^6}\cdot \frac{a^3}{b^2}-\frac{3}{2^8}\cdot \frac{a^5}{b^4}+\cdots$$
 \ \ $$\mu_2(f_{en}(x))=b^2+\frac{3}{4}a^2+\frac{9}{2^7}\cdot \frac{a^4}{b^2}+\frac{51}{2^9}\cdot \frac{a^6}{b^4}+\cdots$$
$$\mu_3(f_{en}(x))=\frac{3}{2}ab^2+\frac{3}{2}a^3+\frac{111}{2^{10}}\cdot \frac{a^5}{b^2}+\cdots ,$$
and then the parameters of the $f_{en}(x)$ in form:
     $$a=\frac{2}{3}\gamma_1n^{-\frac{1}{2}}-\frac{2}{3^3}\gamma_1^3n^{-\frac{3}{2}}+O(\gamma_1^5n^{-\frac{5}{2}}),$$
     $$b^2=1-\frac{1}{3}\gamma_1^2n^{-1}+\frac{15}{2^33^2}\gamma_1^4n^{-2}+O(\gamma_1^6n^{-3}),$$
     $$c=-\frac{1}{2}\gamma_1n^{-\frac{1}{2}}+\frac{3}{2^33^2}\gamma_1^3n^{-\frac{3}{2}}+O(\gamma_1^5n^{-\frac{5}{2}}),$$
when $n^{-\frac{1}{2}}\gamma_1<1$.

For better accuracy of approximation of $f(x)_{\Sigma}$ by the $f_{en}(x)$, in specific cases we can introduce free parameters $a_1, b_1, c_1$:
$$a=\frac{2}{3}\theta+a_1\theta^2, \ \ \ b^2=1+b_1\theta^2, \ \ \ c=-\frac{1}{2}\theta+c_1\theta^2,$$
where $\theta=n^{-1/2}\gamma_1<1.$ and take
into account the last term of Edgeworth expansion (20), which is $\sim O(n^{-1})$.
As one can see, it is impossible to eliminate $R_{4n}(x)$ in general case, but the line $(\frac{1}{4}a_1-\frac{1}{2}b_1-c_1)x+\frac{1}{2}b_1$, which is part of the $R_{4n}(x)$, can decrease $R_{4n}(x) \sim O(n^{-1})$ in any region.

The property (approximation of finite number of asymmetric random variables) allows to interpret the $f_{en}(x)$ analogously with the variant of the popular interpretation of the normal distribution as measurement error distribution. $\xi$ is elementary measuring ($\xi-m$ is elementary error). Asymmetry of the $\xi$ may be caused by either object of the measuring or by device of the measuring. If in practice the finite number, $n$, of essential factors (but not infinity number of infinitesimal) affects on result of the measuring, $\chi$, that asymmetry of the $\xi$ is not terminated and the distribution of $\chi$ is asymmetric. The $f_{en}(x)$ approximate this asymmetric distribution with accuracy $\sim 1/n$ because can not take into account many other characteristics of the $\xi$. (Infinite $n$ terminates these characteristics together with the skewness)$ ^($\footnote {Authors can't point out physical process which the $f_{en}(x)$ describes exactly.}.

It is obvious, that not only the $f_{en}(x)$ has this property. Interpretation of the parameters and simplicity of use of the normal and exponential distributions is inherited by $f_{en}(x)$.

For illustration, we shall examine a concrete example of the study of the cascade maximum depths in the atmosphere using the information from [15]. $P(t_m)$ is the distribution of the depth of maximum, $t_m$, of 500 simulated cascades generated by protons with energy 10$^{17}$ eV at depth $t$ (g/cm$^2$) in the atmosphere, shown as the histogram presented in fig. 3.

One can draw by hand a smooth curve approximating the histogram (see the dashed curve in fig. 3). It is a right-asymmetric unimodal distribution. Let us drop perpendicular from the curve maximum (670, 35) to the abscissa and obtain the value of the mode $c=670 g/cm^2$. Then draw the horizontal line approximately at the level 1/3 (e$\approx$3) of the maximum of the curve and mark the values of $W_l$=58 and $W_r$=94. Then (15, 16) yield $a$=94-58=36, $b=(58\cdot94/2)^{1/2}=52.2$; so the analytical approximation by form (1) can be written: $ 35exp(-(t-670)^2/(36(t-670)+2\cdot 52.2^2))$.

The distribution does not contradict to the physical understanding of the problem and the complete description in [15]. One can test accuracy and acceptability of the formula by the same way as for using the normal distribution. For example, it is very simple to compare the mean value of the approximation with that calculated directly, 699.63 g/cm$^2$. For this, one can calculate the $z$ value defined by (5): $z=(2\cdot 52.2/36)^2=8.41$. Then, using the values of $z$ and $a$, from table 1, one can obtain $\tilde x =0.733\cdot 36=27.83$; so account of shift (c=670) in (10) yields the mean value of the approximation: $\tilde X =670+27.83=697.8 g/cm^2$. (Such good fit for a hand drawing is accidental). To obtain the probability density function $P(t_{m})$, it is only necessary to know the normalization factor. It can be obtained from table 1 by using $z$ and $a$: $A(a,b)=0.260/36=7.2\cdot 10^{-3}$.

This example illustrates a qualitative analysis and using the table 1. The distribution may be used as the first approximation for the correcting process. Regular methods considering specifics of the initial data are described below. The efficiency of application of particular properties of the distribution depends on many things: on the available set of the data under conditions of a concrete problem and even on the form of representation of this information.

\section{Methods of creation of approximation}
Although it is obvious that $f_{en}(x)$ can be determined by the first three moments and the general methods are known, we focus our attention to the problems with the incomplete or poor-quality data. It is a typical case in the investigation of complex objects, for example for UHE cascades. Since the cascades are rare, difficulty registered and one can see only part of the cascade (in particular, it is longer than deep of atmosphere), input data may have errors about 30\% and may be incomplete. For example, some details may be seen in [11, 16]. Exclusions are so-called 'golden events' which are used as etalons. Thus, for accounting asymmetry use a third moment may be incorrect (or in any case is unreliable). The above properties of $f_{en}(x)$ make it feasible to create regular methods of approximation (in contrast to the by-hand method described in the previous section) accounting for asymmetry without using third or higher moments.

Method A. Approximation with conservation of mean, variance and ratio of the variance's parts.

If the mean and variance are calculated, obtaining the parameters of the exponential and normal distributions is reduced to the comparison of the calculated variance with the variance of a standard form (for determination of the scaling parameter) and to the comparison of the calculated mean with the mean value of the standard form (for determination of the parameter shift). Knowledge of the values of the mean and variance is not enough to determine the parameters of $f_{en}(x)$, since the standard form of this distribution is not unique but is one-parametrical series depending on a scaling-invariant parameter, which defines type of the distribution and may be taken as measure of asymmetry.
 Therefore, if the value of a scaling-invariant parameter is found first, the problem of obtaining the values of the parameters of $f_{en}(x)$ will be reduced, in essence, to the same well-known procedures that are used when dealing with the well-known neighboring distributions.

In this connection, one can use the measure of asymmetry introduced by (8), since the value of $R^2$ can be found using the initial data. This value allows one to find (for example, with the help of fig. 1 or table 1) the parameter $z$ of the distribution, i.e. to solve the problem, remaining within the framework of habitual procedures of work with the well-known distributions.

Formula (8) specifies an intuitive estimate of $R^2$ (the so-called method of substitution). Taking into account continuity of all functions under consideration, the estimate will obviously be consistent. Its properties will need to be studied in more detail before wide practical use of $f_{en}(x)$.

So, using the initial data, 3 numbers are calculated: the mean value, $\tilde X$, the left and the right parts of the variance, $\sigma_l^2, \ \sigma_r^2$. Then value of $R^2$ (8) determines the value of parameter $z$ with the help of table 1. The value of $z$ determines both values of variance, $[\sigma ^2/a^2]$, and mathematical expectation, $[\tilde x /a]$, of the standard form of $f_{en}(x)$, \ $[a=1, \ c=0]$. A comparison of the calculated variance with its obtained value, $[\sigma ^2/a^2]$, determines the parameter $a=\sqrt {(\sigma_l^2+\sigma_r^2)/ [\sigma ^2/a^2]}$. Then similarly by taking into account shift, we obtain: $c=\tilde x -[\tilde x /a]\sqrt {( \sigma_l^2+ \sigma_r^2)/ [\sigma ^2/a^2]}$.
To complete the discussion, the parameter $b$ is obtained from (5); and the parameter $A$ is obtained from (3) or with the help of table 1. $ ^($ \footnote {These 3 numbers ($\tilde X, \sigma^2_l, \sigma^2_r$) are representation of $f_{en}(x)$. The numbers ($\tilde X, \sigma_l, \sigma_r$) are convenient for graphical representation of the asymmetrical distribution function, which is similar to traditional representation results of a measuring by mean value and error bars.} The distribution $f_{en}(x)$ with the parameters calculated in this way will be constructed to have the mean, variance and ratio of the variance components of the data. (The numerical example is presented in footnote 3 of section 5).

This method is suitable for various presentations of the initial data; for example, as histograms (interpreted as density of the distribution function, or without such interpretation and considered as values of some function obtained by simulation of a phenomenon), or as a set of results of direct measurements of some physical characteristics; since the method only requires it be possible to calculate the mean and the left and right components of the variance. For direct measurements the stage of creating histograms can be omitted in data processing. The method can be used for poor statistics (say less then 50) when it is difficult to obtain a reliable histogram. The procedure of the method can be easily inserted in a popular procedure of previous data processing (calculation of the mean and dispersion). At last, the method allows checking the applicability of (1) by testing the equation (9) for the asymmetry.

Method B. Approximation by two pairs of parameters.

Let us study a way of approximation based on a specific property of form (1) using the linear (at $x$) form of the denominator in the exponent:
\begin{equation}
a(x-c)+2b^2 =(x-c)^2 /ln(A/f(x)).
\end{equation}
It can be seen that the approximation $f(x)$ (in particular, the histogram), at the known position of the maximum of (1), i.e. at the known value of pair parameters $A$ and $c$, is reduced to the linear approximation of the computable right part of (24).

The pair $(A,c)$ determines the position of a curve (1) on the coordinate plane and its central part, but actually the form and scaling is determined by the pair $(a,b)$ (see (5), (15), (16)). The initial data and their processing often differ in the central and in the peripheral parts of the distribution, because different processes dominate in the central part and at the periphery. Both for physical and statistical reasons, the central part is usually less subject to fluctuations than the periphery. In the case of small fluctuations in the central part, it is possible to obtain enough accuracy the values of $A$ and $c$ by a fitting maximum of the region ($e^{-x^2}=1-x^2+x^4/2- ...$) to a parabola which approximate, say, 5 points at the region of maximum. It is possible (and frequently useful) to exclude some of these points from the subsequent calculation (24) when obtaining $a$ and $b$. An approximation constructed in this way is based on a concept of a mode (instead of the mean as in method A) as a more natural and robust characteristic.

Fig. 4 illustrates method B for a simulated individual shower. The shower was generated by a proton with energy $10^{19}$eV and zenith angle $\theta=44.4^{\circ}$. The circles show the input values $N(t)$, the solid line is for approximation (1) and the squares are for the values of the right part of (24). The area near the maximum is presented for more detail in the inset. The parabola determining $t_m$ and $N(t_m)$ is given by the dashed line. For the linear (at $t$) approximation determining the values of $a$ and $b$
$$a(t -t_m)+2b^2 =(t -t_m)^2 /ln(N(t_m)/N(t))$$
 the ordinate axis with natural more detailed ($\approx $ 4 order) scale is located to the right. From the figure it can be obtained that $N_m=7.39 \cdot 10^9, \ \ t_m=740; \ \ a \approx (1.43\cdot 10^5-7.8\cdot 10^4)/1000=65, \ 2b^2\approx 1.43\cdot10 ^5-(1400-740)=1\cdot 10^5$. So we have:
$$N(t) =7.39\cdot 10^9 exp(-(t-740)^2 /(65(t-740)+1\cdot 10^5)).$$
This method is suitable for truncated data, i.e. in the cases when the tail (or nose) of a distribution is unknown, or for incomplete data with "holes" (for example, when an intermediate part of whole diapason is inaccessible for measuring or a histogram columns are lost during data transmission) since the method requires calculation of the maximum and the linear approximation. From fig. 4 it can be easily estimated how the removal of some points will affect the approximation, so one can judge the quality of the approximation.

Method B is fitting method essentially and based on the robust characteristics of the data. For the case of sufficient statistics the method has an advantage over method A, which can be used for complete data only.

Concrete conditions may lead to construction of methods intermediate between methods A and B. For example, if the mode and mean can be determined quite reliably while the parts of variance seems unreliable, the K. Pearson skewness $(\tilde x -c)/\sigma $ \ [14, 1946] based on the mode can be used to construct $f_{en}(x)$.

Results of the methods can be used as starting point for refinement by fitting procedures.

\section{Some general problems application for UHE cascades}

Ultra-high-energy (UHE) events are very rare, so this fact leads to two consequences: a limitation of using the usual probability theory apparatus and the highly important role of the Monte-Carlo method. Since the phenomenon is complex, the central processor unit time required would rise in proportional to primary energy and a single shower could take well over a century at $10^{20}$ eV ([4] 1997). Practical modeling is carried out by a hybrid method that introduces unphysical fluctuations. Therefore when analyzing the data, it is important to pay attention to the quality input data for each step of the multi-step data processing. In this section we discuss in brief some general problems of the application of form (1) for the UHE cascades. (For some more details see [5]).

The results of Monte Carlo simulation of air showers generated by protons with energies of 10$^{19}$, 10$^{20}$ and 10$^{21}$ eV for zenith angles $\theta$=0$^{\circ}$, 24.6$^{\circ}$, 44.6$^{\circ}$, 60$^{\circ}$ are used. Calculations were carried out in the framework of the quark-gluon-string model [17]. The hybrid method [18] was used to estimate both the mean values and the standard deviations of the parameters under consideration. The method enables accounting for fluctuations of the number and the location of interactions in the atmosphere and energy release of primary protons, using the Monte-Carlo procedure, while the development of cascades from numerous charged pions is considered in an average of step-by-step approach [19]. For energies higher than the threshold value of 10$^{16}$ - 10$^{17}$ eV (depending on the position in the atmosphere), the propagation of electrons and photons was simulated with account for Landau-Pomeranchuk-Migdal (LPM) effect suppressing of bremsstrahlung and pair production [20]. (Details of the simulation for Yakutsk experiment can be seen in [21]). The results obtained were presented as functions at the atmospheric depth $t$ with step $\Delta t$ =50 g/cm$^2$. The technical quality of the simulation results was rather high; local unevenness and unphysical fluctuations are rare, their deviations are, as a rule, less than the natural cascade fluctuations.

For approximation of UHE cascades use is made of various formulae.
A well-known Gaisser-Hillas formula was introduced by [22] to describe the mean particle number $N(E_0 ,t)$ in a vertical $(\theta =0^\circ )$ shower generated by a proton of energy $E_0\ge 10^{15}$ eV at depth $t_0$:
\begin{equation}
N(E_{0},t^{\prime} )=N_{0}(E_{0}/c)exp(t_m^{\prime} )\left(\frac{t^{\prime}}{t_m^{\prime}}\right)^{t_m^{\prime}}e^{-t^{\prime}}
\end{equation}
($t^{\prime}$ is measured in units $\lambda$=70 g/cm$^2$ from the generation depth $t_0$; the depth of the maximum is $t^{\prime}_m$=0.51 ln$(E_0/c)-1$ and $N_0$=0.045, $c$=0.074 GeV). This formula was used later to approximate individual cascades, assuming it to comprise four parameters: $t_m, N_m $ for the position of the maximum and $t_0,\  \lambda$ are free parameters.

Other functional forms have also been used; e.g., two parts of Gaussian functions with two different widths $\sigma _l$ and $\sigma_2$ is used for $t<t_m$ and $t>t_m$, respectively, [23]; or other forms [16], for example, above mentioned parameterization using the width of the distribution [13]. In all cases the approximation accuracy should be much less than physical fluctuations, since frequently the fluctuations are the object of interest when UHE events are investigated.

We use the approximation of giant cascades developed in [2]. The number of each sort of particles (electrons, muons and their sums) in each individual cascade was approximated by form (1) as a function of $t \ge 0.5t_m$ with an accuracy about $1.5\% $. In particular, for the total number, $N^i (t)$, of particles at depth $t$ in an individual cascade number $i$
\begin{equation}
N^i (t)=N_m^i exp(-(t-t_m^i)/(a^i (t-t_m^i )+(2b^i)^2)) \ \ \mbox {at} \ \ t \ge 0.5t^i_m,\,\, i=1 \div 500,
\end{equation}
where $t_m^i, N(t_m^i)$ are the maximum of the cascade, $a^i$ is the parameter of asymmetry, and $b^i$ can be thought a breadth parameter. Approximation (26) has been shown in [2] to be stable. The form describes well both the mean and strongly fluctuating individual cascades.

We describe each individual cascade by 4 parameters; an individual cascade is a concrete realization of 4 random variables with probability density denoted as $P(t_m), P(N_m), P(a)$ and $P(b)$ which describe the entire ensemble of cascades. This approach leads to the definition of a mean cascade as a cascade with parameters each of them mathematical expectation. Such definition is a natural generalization of the traditional definition of a mean cascade where the number of particles at the depth $t$ takes one its expectation value.

The function $P(t_m)$ shows a narrow peak at the left side from the mean value, $\tilde t_m $, and a wide range ($\approx 6\sigma$) of the parameter with approximately exponential asymptotic attenuation. The function $P(N_m)$ has the parameter range $\approx 5\sigma$, the mean shifted to the right and the maximum to the right from the mean, with a peak wider than that of $P(t_m)$. The function $P(a)$ has the parameter range $\approx 4\sigma$, its form is similar  to $P(N_m)$. The function $P(b)$ is similar to $P(t_m)$, the mean value and the maximum are shifted to the left. The distribution functions describe the model results in full detail in a compact form (for example see fig. 6 below).
 The functions are very sensitive: small changes of the simulation (whose influence is difficult to see) are noticeable in the distribution functions.

Since cascade processes dominate in the UHE cascades, asymmetry of the cascades is not so large ($z>>4$, see for the example fig. 4), that $\int _{0}^{\infty} {N^i(t)dt} \approx \sqrt{2\pi}N_m^i b^I,$ see [2] for details. It is useful to define a new parameter $s^i=\sqrt{2\pi}N_m^i b^I,$ which fluctuates less than $N_m$ and may be used for primary energy determination.
Analysis of the distribution functions reveals showers with superfluous characteristics. Special investigation of $\approx$0.5\% outlying showers shows some shortcomings of the model and simulation techniques and a way to correct them. The parameter $s$ is the most useful for revealing the unphysical fluctuations. Thus, the known interpretation of the parameters enables checking the simulation.

In cases high quality of the data makes it feasible to apply both A and B methods. However, for higher energies and smaller angles, as well as for consideration of some characteristics of cascades (for example, muon flux) the data will be incomplete. Vertical cascades of higher energies poorly reach a maximum at the sea level, so only half of the distribution can be observed. Besides that it is necessary to take into account restricting outer edges of (26) ($t\ge 0.5t_m$) and the increase fluctuations at the cascade's peripheries. When both methods are applicable, they give practically the same results: the difference in $t_m$ is, as a rule, less than 3g/cm$^2$. (For comparison, accuracy of the parameter at experimental measuring is $\approx$ 10 g/cm$^2$ in the best special cases [11]).

\section{Distribution function of the depth of cascade maximum}

The shower maximum position, $t_m, N_m$, is measured directly [11]. The characteristic is used for estimation of primary energy and mass composition of primary cosmic ray flux. The cosmic ray composition is studied by comparing the observed values with predications from simulations for different nuclei. The change of the observed $t_m$ distribution is used to derive estimates of the change in primary composition. At a certain energy the average value $\tilde t_m$ of primary mass $A$, is related to mean $\tilde t_m$ of the proton.

Physical basis for such way investigation is domination of proton component ($\approx 80\%$) in flux of cosmic rays and known differences between cascades generated by nuclei and photons and cascades generated by protons. Then briefly speaking, due to larger multiplicity in firsts interactions of the showers generated by nuclei develop faster and less fluctuate than proton ones. Therefore (at the same energy) the distribution function of $t_m$ from nuclear primary is shifted at less depth ($\approx 70g/cm^2$) and has less width (as think the authors, the skewness must be less too). But cascades from photon primary due to less multiplicity difference from proton ones to the opposite side, i.e. its depths are larger and distribution functions are wider.

Thus, $t_m$ of proton generated showers is basically characteristic for energy and mass composition estimates. For these estimations are used the showers which estimations of uncertainness $t_m$ and total energy are smaller then 40$g/cm^2$ and 20\%, respectively [12].

The depth of cascade maximum is determined by two physical processes: by the depth of the first interaction (with a known exponential distribution) and by fluctuating cascade development (from the first interaction point) which resembles the central part of the normal distribution. The distribution function of the depth is not a convolution of the exponential and normal distributions, but rather the function must be determined on the half-line, has a maximum region similar to the normal distribution, and has an exponential attenuation. Therefore, it is natural to use $f_{en}(x)$ as an approximation of the probability density of the cascade maximum depth, $P(t_m)$. As will be seen below, the first process dominate since the asymmetry is significant and invariant parameter of the approximation $z<4$.

To estimate $P(t_m)$, we use described in previous section simulated cascades of energies of 10$^{19}$, 10$^{20}$ and 10$^{21}$ eV for zenith angles $\theta$=0$^{\circ}$, 24.6$^{\circ}$, 44.4$^{\circ}$, 60$^{\circ}$. The cascades are first approximated by $f_{en}(x)$ using method B. This representation yields $t_m$ for each of the simulated cascades and the total set of $\{t_m\}$ is input data for the step of obtaining a distribution function. Note for the second step the quality of the data may be worse than input data for the previous step, because an error of each $t_m$ can lead to unphysical fluctuations: local groups and rarefactions. The statistics may be poor (especially for natural events), but the data have not any 'holes' and cut of the tail. Therefore for the step we choose method A just to the initial set $\{t_m\}$. (As recommended by section 4, histogram of $\{t_m\}$ is not required)$ ^($\footnote {For example, take case, $E_p=10^{19}$eV with zenith angle $\theta$ = 44.4$^\circ$. In accordance with method A (section 3), calculation gives the mean value $\tilde t_m $ = 785, the left and right components of the variance: $s_l^2$ = 514 and $s_r^2$ = 1255. We calculate $R^2=s_l^2/s_r^2$ = 0.41 and, using fig.1 or 5$\rightarrow$1 lines of table 1, find the value $z$ = 1.3. For this $z$, table 1 gives: $A(a,b)\vert a\vert$ = 0.568, $\tilde x/a$ = 0.841, $\sigma ^2/a^2$ = 1.18. Then it is consistently calculated that $a^2$ = (514+1255)/1.18 = 1499, $(2b)^2$ = 1.3/1499, ($a$ = 38.7, $b$ = 22.1), $A(a,b)$ = 0.568/38.7 = 0.0147, $\tilde x  = 0.841\cdot 38.7$ = 32.5, $c$ = 785-32.5 = 752.5.}.

For the case $E_p=10^{19}$eV, $\theta$=44.4$^{\circ}$, the probability density $P(t_m)$ with the parameters $A(a,b)$ = 0.0147, $a$ = 38.7, $b$ = 22.1, $c$ = 752.5 is presented in fig. 5 by solid curve. Though the distribution is obtained without any histogram, for visual illustration of approximation method A quality there is one histogram of {$t_m$} which is given by solid straight-lines.
Dotted histogram is experimental data [10] shifted on 100$g/cm^2$ to the great depth. The discussion of the experimental data will be below. The isolated fragment is representation of the distribution $P(t_m)$ through mean value and components of variance. The representation was discussed in connection with method A and mentioned in footnote 2. Note that accuracy of mean value may be presented by diameter of the point which denotes the mean value. (The representation gives ability clear to see the curve of the $f_{en}(x)$ distribution as $-\sigma \cdot +\sigma$ to see normal distribution).

The construction of the probability density $P(t_m)$ was carried out, in essence, in an integral approach, avoiding a histogram, therefore, it is natural to use the Kolmogorov's test. The goodness of fit between the empirical distribution function $F^*(t_m)$ (sum polygon) and the obtained distribution function $F(t_m)$

$$F(t_m) = \int_ {-\infty}^{t_m}{P(t, a=38.7, b=22.1, c=752.5)dt}, $$
yields: $\lambda = \sqrt{n}$ max$\vert F(t_m) -F^* (t_m)\vert$ = 0.59 and, accordingly, the probability of the observed discrepancy, $P(\lambda )\approx$0.88, may easily be explained by statistical fluctuations.
A similar consideration was made for other energies: $10^{20}(eV)\Rightarrow
P(\lambda)\approx 0.71, \ 10^{21}(eV)\Rightarrow P(\lambda)\approx 0.72$.

As soon as statistics (500) permits to check the A method, let us verify the results by $\chi^2$-test. We use the histogram with columns presenting probability equal to $1/k$, where $k$ is the total number of the columns and let $k=20$ (example for $E_p=10^{19} eV$ see in fig. 5). Pearson's $\chi^2$ test of the simple hypothesis ($a=38.7,\ \ b=22.1,\ \ c=752.4$)\ \ against an alternative yields $P_{19}(\chi^2)\approx 45\%$ (for 19 degrees of freedom), then for composite hypothesis $\chi^2$-minimum method yields $a=41.5, \ \ b=17.9, \ \ c=749.5$ and $P_{16}(\chi^2)\approx 70\%$. The results for simple hypothesis are listed in the left, but ones for the composite hypothesis are listed in the right side of table 2.
When analyzing the table, we take into account that the results for $10^{20}$ are less accurate then others.

Due to UHE events are rare; comparison with experimental data is difficult. We can take distribution of $t_m$ for natural all-energy, all-zenith angle and all-kind primaries [10], only. As it will be shown below (see fig. 6) we can neglect the angle dependence. Approximately 20\% of the mixture is iron which has $t_m$ less then ones of proton $\approx$ 85 $gcm^{-2}$. The energy dependence may be taken into account partly and only roughly. Energy spectrum of flux in the region of so-called 'ankle' ($\approx 2\cdot 10^{18}$eV, see fig. 7) is complex (but about $E^{-3}$); in any case the majority of the mixture consists the cascades of low energies ($\approx 10^{18}$ eV). Increasing average $t_m$ with increasing energy at one order (so-called 'elongation rate') is about 71 $gcm^{-2}$/decade at 10$^{18}$ eV [11]. Thus, for comparison with proton primary $t_m$ at energy $10^{19}$ eV we must to shift the taken data to large dept at the least 71+0.2 $\cdot$ 85=87 (g/cm $^2$). To make visual comparison more convenient we shift the experimental data at 100 g/cm $^2$ in fig. 5. It is needed to have in view that both energy dependence and non-proton primaries wide the distribution significantly. So fig. 5 shows general agreement the simulated and experimental data. Remembering the descriptive ability of $f_{en}(x)$ (fig. 2) and quality data authors think that $f_{en}(x)$ may be used to investigate distribution function of the depth of cascade maximum.

The functions, $P(t_m)$, were obtained in above discussed way for all simulated cases primary protons: energies of 10$^{19}$, 10$^{20}$ and 10$^{21}$ eV for zenith angles $\theta$=0$^{\circ}$, 24.6$^{\circ}$, 44.4$^{\circ}$, 60$^{\circ}$. The full description of the distribution functions of $t_m$ is given by fig. 6 into compact form through mean values and components of variances. The mean values of $t_m$ are independent of angle, practically, and the elongation rate of proton fraction is seen clearly ($\approx 55 g/cm^2$), that is the same as [11]. The variance is decreasing at increasing the energy ($\approx 25\% $ at all diapason). The position is natural and agrees with previous results for 45$^{\circ}$ [15]. Note, together with decreasing the variance one can see some increasing skewness of the distribution, but we think the conclusion may be unreliable (for detail see [5 (2007)]). Taking into account the quality of the data now, we can see scaling dependence of the distribution form at energy, only. The width of the band containing 50\% of the probability is approximately 50g/cm$^2$ (see also fig. 5). Thus, to define the primary energy, the parameter $t^i_m$ is inferior to $s^i = \sqrt{2\pi}N^i_mb^i$.

Formal apparatus of distribution functions gives ability to obtain not only semi-quality estimations (as above), but correct quantity estimations. To illustrate technical work with cumulative distribution presented by table 1 we will obtain the region of overlap the $P(t_m)$ for $\theta$=44.4$^{\circ}$ at energy $10^{19}$eV with one at energy $10^{20}$eV. The parameters of the $P(t_m)$ take from left side of table 2. The $P(t_m)$ at $10^{20}$ eV has left bound $t_m>c-2\cdot b^2/a=807-2\cdot 8.5^2/40=803.4 (g/cm^2)$. At the depth cumulative distribution (19) for $10^{19}$ eV $F(t_m, 38.7,22.1,752.5)=0.75$. $ ^($\footnote {From (19) we have $y=(803.4-752.5)/38.7=1.32, z=(2\cdot 22.1/38.7)^2=1.3$. Then from table 1, by interpolation we have $F(1.32, 1.3)=0.75$}. Thus, proton generated showers which energy differ more then in ten times can have the same depth of maximum with probability 0.25. To clear the problem, for example, one can create function of maximum probability at plane $t_m$, - primary $E$, easily.

\section{Discussion}

As noted above, report [11] studies the cosmic ray composition in different energy ranges by comparing the observed average $t_m$ with predictions from air shower simulations for different nuclei. The general situation is presented in figure 3 of the report. We added our results of distribution function of $t_m$ for proton generated showers: top straight line segment at $10^{19}$ eV with arrows denoted $\sigma _l$, $\sigma _r$ and lower conditional boundary for value of $t_m$. See fig. 7. In our opinion, the difficulties in mass-spectra estimation due to not differences mean results of various models (as it was some years ago [15]), but wide band of physical fluctuations, i.e. wide width of distribution function. Therefore we hope the formal apparatus of distribution functions will be useful for the problem.

At the highest energies, due to relict suppression of nuclear flux [1], it would be expected a significant proportion (10\%$\div$ 50\%) of the spectrum of cosmic rays would be photons [12]. The $t_m$ of photon generated shower is much greater then their nuclear counterparts. This is due to both causes: much lower multiplicity in particle production in electromagnetic - dominated photon showers than in the hadronic present for nuclear primaries and suppression of Bete-Heitler pair production by LPM effect [20], which is not important for other cosmic ray primaries. The influence of LPM effect on the form of cascade curve is well-known from electromagnetic cascade theory. Thus, both the results and $t_m$ can be used to discriminate between photon and nucleonic UHE primaries. More exactly estimation of the proportion may be carried out with help of formal apparatus of distribution functions (not only visual comparison of histograms). The increase of the statistic UHE events will give all needed for the ability.

\section{Summary}

The asymmetric form $f_{en}(x)=Aexp(-(x-c)^2/(a(x-c)+2b^2))$ inherits the convenience of interpretation of its parameters from exponential and normal distributions. The methods offer complementary options to deal with incomplete and low-quality data, while remaining comfortable with the usual procedures for normal distribution. In particular, application of the presented apparatus for UHE-cascades enables to obtain the correct description of individual cascades generated by UHE protons and to investigate the distribution function of the depth of their maximum.

\bf{Acknowledgements}

\rm
The authors are grateful to professors V. Yr. Korolev and L.G. Dedenko for their favors of the work and G.F. Fedorova for performing calculations.

The work is maintained by the 03-02-16290 grant of the Russian Federal Property Fund and by the 03-51-5112 grant of INTAS.

\vskip 1cm

REFERENCES

 [1] \ \ K. Greizen. Phys. Rev. Lett. 2, (1966) 748.

 \ \ \ \ \ \ G.T. Zatsepin, V.A. Kuzmin. JETF Lett. v. 4, (1966) 53.

 [2] \ \ A.A. Kirillov, I.A. Kirillov. Astropart. Phys. 19, (2003) 101.

 [3] \ \ M. Roth, for the Auger Collaboration. 30th. Int. Cosmic Ray Conf., Proceedings-Pre-Conf. Edit. HE.1.4.A, pap. [313].

 [4] \ \ A.M. Hillas. Nucl. Phys. B (Proc. Suppl.) 52B, (1997) 29.

 [5] \ \ L.G. Dedenko, A.A. Kirillov, et al. Proc. 28th. Int. Cosmic Ray Conf., vol. 2. (2003) p. 531.

\ \ \ \ \ \ A.A. Kirillov, I.A. Kirillov. Proc. 28th. Int. Cosmic Ray Conf.,
vol. 2. (2003) p. 535.

\ \ \ \ \ \ A.A. Kirillov, I.A. Kirillov, G.P. Shoziyoev. Proc. 29th. Int.
Cosmic Ray Conf., vol. 7. (2005) p. 259.

\ \ \ \ \ \ O.V. Bondartsova, A.A. Kirillov, Bulletin of the Russian Academy of Sciences. Physics. (Izvestia Rossiiskoy Akademii Nauk. Seriya Fizicheskaya, 2007, tom 71, N. 4, c. 527-529, in Russian).

 [6] \ a. \ O. E. Barndorff-Nielsen. Proc. R. Soc. Lond. A. 353, (1977) 401.

 \ \ \ \ \ b. \ O. E. Barndorff-Nielsen, Ch. Halgreen. Z. Wahrscheinlichkeitstheoric verw. Gebiete 38, \ (1977) 309.

 \ \ \ \ \ \ c. \ O. E. Barndorff-Nielsen, D.R. Cox. Asymptotic Techniques for Use in Statistics. (1989) London. Chapman and Hall.

 [7] \ B. Jorgensen. Statistical Properties of the Generalized Inverse Gaussian Distribution. Lecture Notes in Statistics, 9, (1982) Springer, New York.

 [8] \ \ I.S. Gradshteyn and I.M. Ryzhik. Table of Integrals, Series, and Products. 4 th edition. (1965) New York: Academic Press.

 [9] \ \ M. Abramowitz, I.A. Stegun. Handbook of Mathematical Functions. (1965) Dover, New York.

[10] \ \ R.U. Abbasi, T. Abu-Zayyad, G. Archbold et al., Astrophys. J., v. 622, part 1 (2005),p. 910

[11] \ \ M. Unger, for the Pierre Auger Collaboration. 30th. Int. Cosmic Ray Conf., Proceedings-Pre-Conf. Edit. HE.1.4.A, pap. [594].

[12] \ \ M. Healy, for the Pierre Auger Collaboration. 30th. Int. Cosmic Ray Conf., Proceedings-Pre-Conf. Edit. HE.1.4.A, pap. [602].

[13] \ \ G. Hughes, for High Resolution Fly's Eye Collaboration. 30th. Int. Cosmic Ray Conf., Proceedings-Pre-Conf. Edit. HE.1.4.A, pap. [712].

[14] \ \ H. Cramer. Random variables and probability distributions. Cambridge Tracts in Mathematics, No. 36, (1937) Cambridge.

\ \ \ \ \ \  H. Cramer. Mathematical methods of statistics. (1946) Stocholm.

[15] \ \ C.I. Pryke. Astropart. Phys. 14, (2001) 319.

[16] \ \ M. Unger, R. Engel, F. Schussler, et al., 30th. Int. Cosmic Ray Conf., Proceedings-Pre-Conf. Edit. HE.1.3.A, pap. [972].

[17] \ \ A.B. Kaidalov, K.A. Ter-Martirosyan, et al. Yad. Fiz. 43 (1986) 1282

[18] \ \ L.G. Dedenko. Can. J. Phys. 1968. v. 46, p. 178.

[19] \ \ L.G. Dedenko. Proc 9th Int. Cosmic Ray Conf.,vol.2, (1966) p.662.

[20] \ \ A.B. Migdal. Phys. Rev. 103 (1956) 1811.

[21] \ \ A.M. Anokhina, L.G. Dedenko, et al., Phys. Rev. D. (1999). V. 60.
P. 033004-1.

[22] \ \ T.K. Gaisser, A.M. Hillas, Proc 15th Int. Cosmic Ray Conf.,vol.8, (1977) p.353.

[23] \ \ M. Giller, A. Kacperczyk, et al., Proc 29th Int. Cosmic Ray Conf.,vol.7, (2005) p.187.

\newpage

\centerline{
Table 1. The values of the functions for calculation of the parameters of $f_{en}(x,a,b)$}

\centerline{and for calculation of the values of $F_{en}(y,z)$.
}

\footnotesize
\begin {tabular}{rrrrrrrrrrr}

$z=$ &.1E-02 &.2E-02 &.5E-02 &.1E-01 &.2E-01 &.5E-01 &.1E+00 &.2E+00 &.5E+00 &.1E+01\\
$A\vert a\vert =$ &.99900 &.99802 &.99509 &.99031 &.98109 &.95556 &.91826 &.85714 &.73233 &.61119\\
$\tilde x /a =$ &.99950 &.99901 &.99757 &.99524 &.99081 &.97891 &.96232 &.93670 &.88952 &.84974\\
$\sigma ^2/a^2-1=$ &.00000 &.00000 &.00001 &.00002 &.00010 &.00061 &.00235 &.00865 &.04304 &.12768\\
$\sigma _l^2/\sigma _r^2=$ &.35914 &.35914 &.35914 &.35915 &.35918 &.35938 &.36006 &.36243 &.37406 &.39829\\
-0.4  &.00000 &.00000 &.00000 &.00000 &.00000 &.00000 &.00000 &.00000 &.00000 &.00303\\
-0.2  &.00000 &.00000 &.00000 &.00000 &.00000 &.00000 &.00000 &.00000 &.00602 &.07670\\

0.0  &.00050 &.00099 &.00246 &.00485 &.00946 &.02221 &.04087 &.07143 &.13384 &.19441\\
0.2  &.18168 &.18208 &.18326 &.18517 &.18880 &.19855 &.21229 &.23383 &.27550 &.31417\\
0.4  &.33001 &.33034 &.33131 &.33287 &.33582 &.34367 &.35449 &.37084 &.39983 &.42329\\
0.6  &.45146 &.45173 &.45252 &.45379 &.45620 &.46257 &.47125 &.48402 &.50486 &.51850\\
0.8  &.55089 &.55111 &.55176 &.55280 &.55477 &.55996 &.56696 &.57707 &.59241 &.59981\\
1.0  &.63230 &.63248 &.63301 &.63386 &.75565 &.63970 &.64538 &.65346 &.66493 &.66841\\
1.2  &.69896 &.69910 &.69954 &.70023 &.70155 &.70500 &.70961 &.71610 &.72479 &.72586\\
1.4  &.75353 &.75365 &.75400 &.75457 &.75599 &.75847 &.76222 &.76745 &.77409 &.77372\\
1.6  &.79820 &.79830 &.79859 &.79906 &.79994 &.80224 &.80530 &.80953 &.81464 &.81346\\
1.8  &.83478 &.83486 &.83510 &.83548 &.83620 &.83809 &.84058 &.84400 &.84797 &.84637\\
2.0  &.86473 &.86480 &.86499 &.86531 &.86590 &.86744 &.86946 &.87224 &.87534 &.87357\\
2.2  &.88925 &.88931 &.88947 &.88972 &.89020 &.89146 &.89312 &.89538 &.89780 &.89602\\
2.6  &.92576 &.92580 &.92591 &.92608 &.92640 &.92724 &.92835 &.92984 &.93134 &.92978\\
3.0  &.95024 &.95026 &.95033 &.95045 &.95067 &.95123 &.95197 &.95296 &.95390 &.95264\\
3.5  &.96982 &.96983 &.96988 &.96995 &.97008 &.97042 &.97086 &.97146 &.97199 &.97110\\
4.0  &.98169 &.98170 &.98173 &.98177 &.98185 &.98206 &.98232 &.98268 &.98299 &.98239\\
5.0  &.99327 &.99327 &.99328 &.99329 &.99332 &.99340 &.99349 &.99362 &.99373 &.99347\\
10.0 &.99995 &.99995 &.99996 &.99995 &.99996 &.99996 &.99995 &.99994 &.99996 &.99996\\
\end {tabular}
\newpage

Continuation of table 1

\begin {tabular}{rrrrrrrrrr}
$z=$ &.2E+01 &.5E+01 &.1E+02 &.2E+02 &.5E+02 &.1E+03 &.2E+03 &.5E+03 &.1E+04\\
$A\vert a\vert =$ &.48380 &.33318 &.24345 &.17518 &.11200 &.07949 &.05631 &.03566 &.02522\\
$\tilde x /a =$ &.81431 &.78149 &.76708 &.75893 &.75368 &.75186 &.75093 &.75037 &.75021\\
$\sigma ^2/a^2-1=$ &.33690 &1.0448 &2.2749 &4.7632 &12.255 &24.752 &49.751 &124.75 &249.73\\
$\sigma _l^2/\sigma _r^2=$ &.44472 &.54088 &.62683 &.70955 &.80079 &.85357 &.89369 &.93123 &.95083\\
-50.0  &.00000 &.00000 &.00000 &.00000 &.00000 &.00000 &.00000 &.00000 &.00038\\
-20.0  &.00000 &.00000 &.00000 &.00000 &.00000 &.00000 &.00061 &.02806 &.09296\\
-10.0  &.00000 &.00000 &.00000 &.00000 &.00007 &.00968 &.05911 &.16908 &.25125\\
-5.0  &.00000 &.00000 &.00000 &.00034 &.04221 &.12401 &.21285 &.30975 &.36337\\
-4.0  &.00000 &.00000 &.00000 &.00551 &.08408 &.17493 &.25838 &.34257 &.38757\\
-3.0  &.00000 &.00000 &.00043 &.03336 &.14743 &.23595 &.30795 &.37649 &.41218\\
-2.5  &.00000 &.00000 &.00514 &.06431 &.18727 &.26980 &.33399 &.39378 &.42460\\
-2.0  &.00000 &.00000 &.02503 &.10990 &.23207 &.30554 &.36072 &.41124 &.43708\\
-1.6  &.00000 &.00218 &.06054 &.15706 &.27102 &.33526 &.38251 &.42532 &.44710\\
-1.2  &.00000 &.02578 &.11696 &.21278 &.31222 &.36579 &.40459 &.43947 &.45715\\
-1.0  &.00000 &.05387 &.15263 &.24335 &.33352 &.38130 &.41572 &.44656 &.46218\\
-0.8  &.00060 &.09397 &.19256 &.27539 &.35517 &.39693 &.42689 &.45367 &.46722\\
-0.6  &.02040 &.14462 &.23597 &.30861 &.37712 &.41267 &.43810 &.46079 &.47226\\
-0.4  &.07842 &.20330 &.28199 &.34272 &.39929 &.42849 &.44933 &.46791 &.47730\\
-0.2  &.16284 &.26715 &.32972 &.37742 &.42161 &.44436 &.46058 &.47504 &.48234\\
0.0 &.25810 &.33341 &.37828 &.41241 &.44400 &.46025 &.47184 &.48217 &.48739\\
0.2  &.35375 &.39971 &.42684 &.44740 &.46639 &.47615 &.48310 &.48930 &.49243\\
0.4  &.44392 &.46418 &.47468 &.48212 &.48871 &.49202 &.49436 &.49643 &.49748\\
0.6  &.52579 &.52545 &.52119 &.51632 &.51088 &.50784 &.50559 &.50356 &.50252\\
0.8  &.59832 &.58261 &.56585 &.54978 &.53286 &.52358 &.51680 &.51067 &.50756\\
1.0  &.66152 &.63511 &.60828 &.58230 &.55457 &.53923 &.52797 &.51778 &.51259\\
1.2  &.71594 &.68272 &.64820 &.61371 &.57595 &.55475 &.53910 &.52488 &.51763\\
1.4  &.76239 &.72544 &.68543 &.64388 &.59696 &.57013 &.55018 &.53196 &.52265\\
1.6  &.80176 &.76342 &.71988 &.67268 &.61753 &.58535 &.56119 &.53903 &.52768\\
1.8  &.83498 &.79692 &.75151 &.70005 &.63763 &.60038 &.57214 &.54608 &.53269\\
2.0  &.86288 &.82627 &.78037 &.72591 &.65722 &.61521 &.58301 &.55311 &.53770\\
2.2  &.88623 &.85184 &.80653 &.75025 &.67625 &.62982 &.59379 &.56011 &.54270\\
2.6  &.92198 &.89309 &.85130 &.79429 &.71255 &.65830 &.61508 &.57405 &.55267\\
3.0  &.94670 &.92356 &.88699 &.83228 &.74634 &.68571 &.63595 &.58788 &.56261\\
3.5  &.96704 &.95026 &.92094 &.87179 &.78487 &.71830 &.66137 &.60498 &.57495\\
4.0  &.97968 &.96794 &.94543 &.90333 &.81921 &.74889 &.68595 &.62184 &.58722\\
5.0  &.99234 &.98697 &.97488 &.94706 &.87566 &.80370 &.73233 &.65476 &.61144\\
10.0 &.99994 &.99988 &.99966 &.99848 &.98797 &.95804 &.89949 &.79746 &.72397\\
20.0 &1.0000 &1.0000 &1.0000 &1.0000 &.99998 &.99949 &.99401 &.95324 &.88665\\
50.0 &1.0000 &1.0000 &1.0000 &1.0000 &1.0000 &1.0000 &1.0000 &.99997 &.99858\\
\end {tabular}

\normalsize

\newpage

\centerline{
Table 2. Results of method A and $\chi ^2$ method (k=20)}

\centerline{
approximation of the probability density of the depth of cascade maximum for $\theta$=44.4$^{\circ}$
}

\centerline{ \ }
\centerline{
\begin {tabular}{l||l|l|l|l|l|l||l|l|l|l|l} \hline \hline
$E_p$(eV) & $A$ & $a$ & $b$ & $c$ & $P(\lambda)$ & $P_{19}(\chi^2)$ & $A$& $a$& $b$& $c$& $P_{16}(\chi^2)$ \\ \hline \hline
$10^{19}$ & 0.0147 & 38.7 & 22.1 & 752.5 & 0.88 & 0.45 & 0.0160 & 41.5 & 17.9 & 749.5 & 0.7\\ \hline
$10^{20}$ & 0.0215 & 40 & 8.5 & 807 & 0.71 & 0.25 & 0.0202 & 37.5 & 12.2 & 811 & 0.23\\ \hline
$10^{21}$ & 0.0255& 27 & 11 & 871 & 0.72 & 0.7 & 0.0289 & 27.8 & 7.8 & 869 & 0.95\\ \hline
\end {tabular} }

\newpage

\begin{figure} [t]
\begin{center}
\epsfig{file=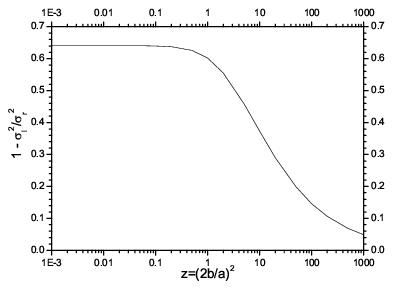, width=15cm}
\end{center}
{   Fig. 1. Dependence of the ratio parts of variance ($1-\sigma _l^2/\sigma _r^2$) on parameter of type of the distribution.}
\end{figure}

\newpage

\begin{figure} [t]
\begin{center}
\epsfig{file=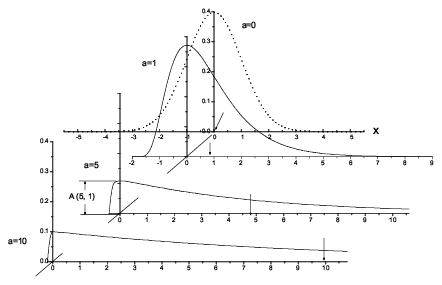, width=15cm}
\end{center}
{   Fig. 2. Example of standard forms of the distribution at $a=0,\  5,\  10;\  b=1$ and $c=0.$ The arrow indicates the mean value for each case. Here increasing $a$ changes $f_{en}(x)$ from normal towards exponential.}
\end{figure}

\newpage

\begin{figure} [t]
\begin{center}
\epsfig{file=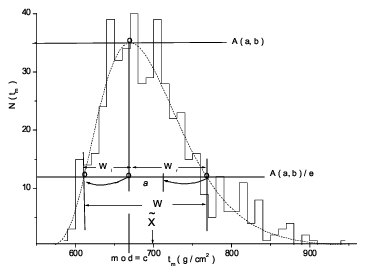, width=15cm}
\end{center}
{   Fig. 3. Interpretation of the parameters and approximation of a histogram by the hand method. For explanation of the construction see the text.}
\end{figure}

\newpage

\begin{figure} [t]
\begin{center}
\epsfig{file=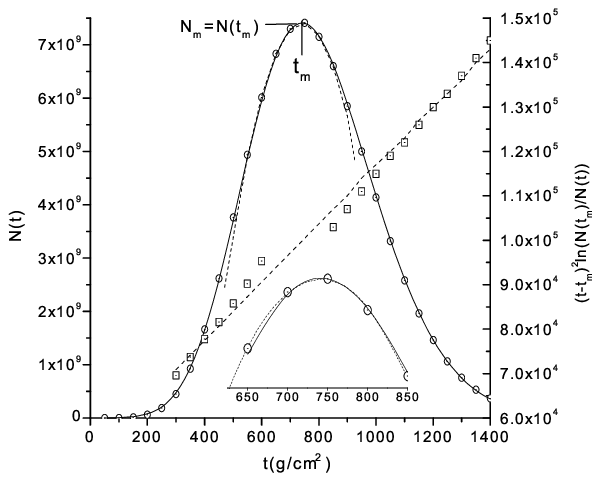, width=15cm}
\end{center}
{   Fig. 4. Approximation of an individual shower by method B. Ordinates for $N(t)$ (circles) are at left, ordinates for $(t-t_m)^2/\ln {(N(t_m)/N(t))}$ (squares) are at right.}
\end{figure}

\newpage

\begin{figure} [t]
\begin{center}
\epsfig{file=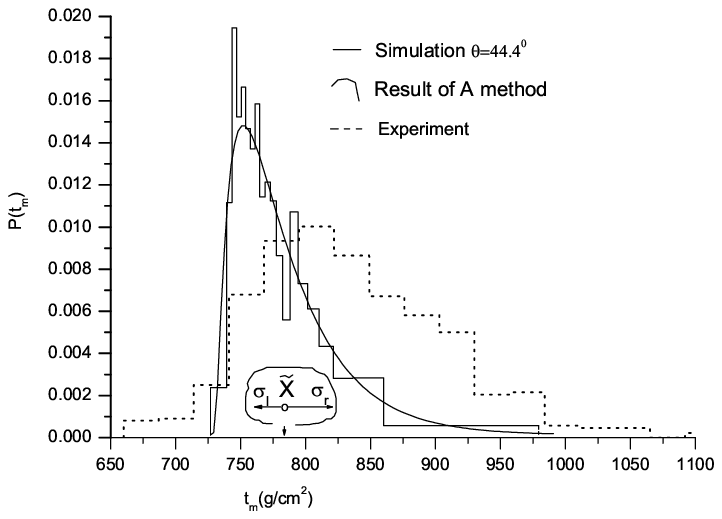, width=15cm}
\end{center}
{   Fig. 5. Probability densities of the depth of UHE proton generated cascade maximum. Isolated fragment is representation of the A method result through mean value and components of variance. Experimental data is all-energy, all-angle and all-kind primary particle mixture, but is shifted on +100 $g/cm^2$. For explanation see text.}
\end{figure}

\newpage

\begin{figure} [t]
\begin{center}
\epsfig{file=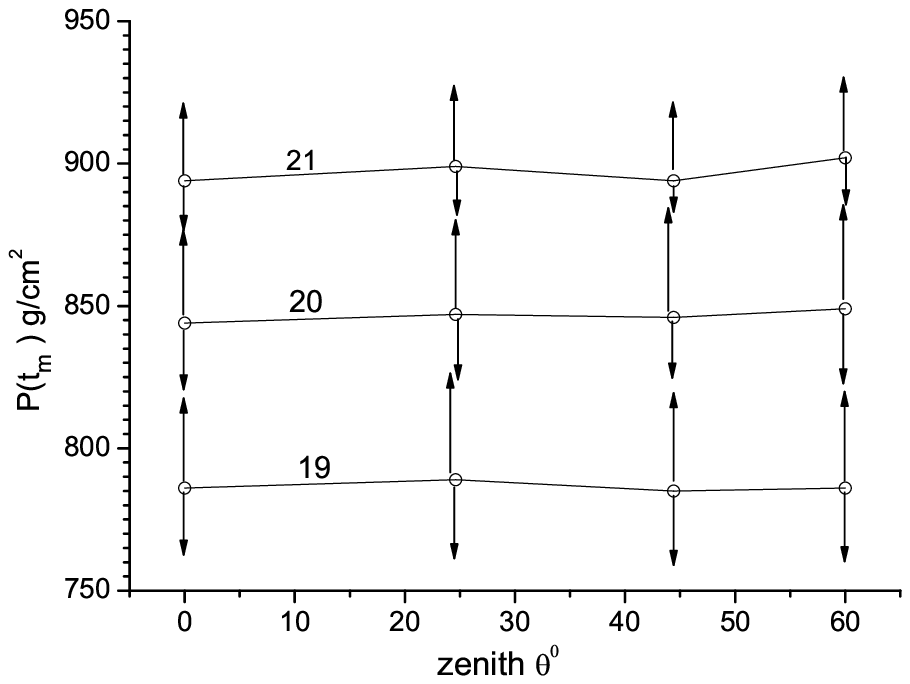, width=15cm}
\end{center}
{   Fig. 6. Distribution functions of depth of simulated cascade maximum generated by protons. Log$_{10}$ energies (eV) are marked at the lines. Diameter points are accuracy of the mean values, arrows are parts of variance. Compare with fig. 5.}
\end{figure}

\newpage

\begin{figure} [t]
\begin{center}
\epsfig{file=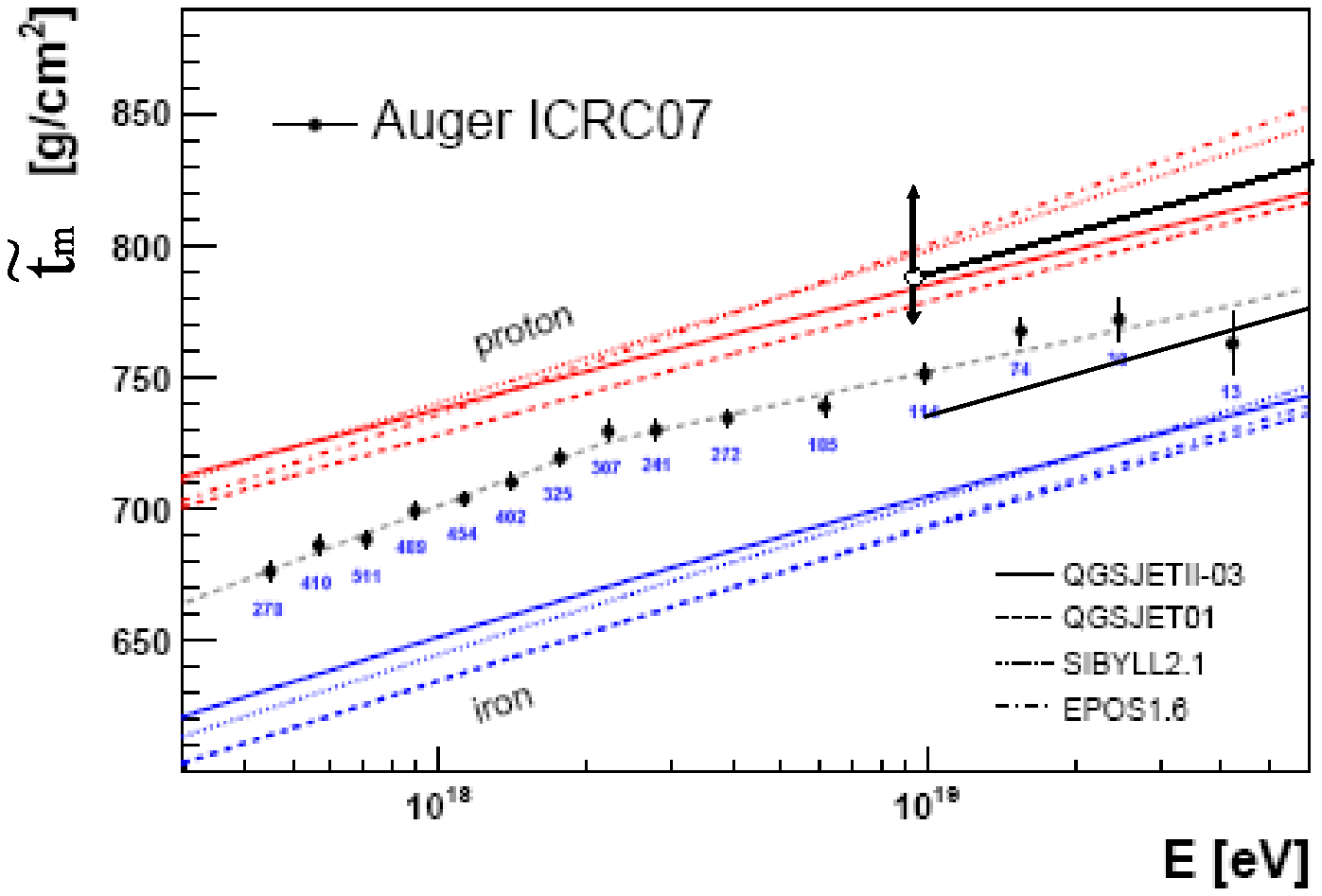, width=15cm}
\end{center}
{   Fig. 7. Mean values of observed $t_m$ compared to predictions from hadronic interaction models. Experimental event numbers are indicated below each data point. Top strait line segment and arrows show distribution function of proton generated $t_m$, lower segment is conditional bound of proton generated $t_m$.}
\end{figure}

\end{document}